\documentclass[conference]{IEEEtran}

\IEEEoverridecommandlockouts
\usepackage{verbatim}
\usepackage{graphicx}
\usepackage{cite}
\usepackage{url}
\usepackage[cmex10]{amsmath}
\usepackage{amssymb}
\usepackage{algorithm}
\usepackage{algpseudocode}
\usepackage{cases}
\usepackage[caption=false,font=footnotesize]{subfig}
\usepackage{color}
\usepackage{cite}
\usepackage{epstopdf}
\usepackage{calc}
\usepackage{array}
\usepackage{algpseudocode}
\usepackage{xparse}
\let\oldState\State
\RenewDocumentCommand{\State}{o}{
  \IfValueTF{#1}{\makeatletter\setcounter{ALG@line}{#1}\addtocounter{ALG@line}{-1}\makeatother}{}%
  \oldState\ignorespaces%
}%

\DeclareGraphicsExtensions{.pdf,.eps,.png,.jpg,.mps}

\begin{document}
\title{Downlink Transceiver Beamforming and Admission Control for Massive MIMO Cognitive Radio Networks}
\author{\IEEEauthorblockN{Shailesh Chaudhari and Danijela Cabric}
\IEEEauthorblockA{Department of Electrical Engineering,
University of California, Los Angeles\\
Email: schaudhari@ucla.edu, danijela@ee.ucla.edu}
\thanks{This work has been supported by the National Science Foundation under grant 1149981.}
}

\maketitle

\begin{abstract}
In this paper, an optimization framework is proposed for joint transceiver beamforming and admission control in massive MIMO cognitive radio networks. The objective of the optimization problem is to support maximum number of secondary users in downlink transmission with constraints on the total power allocated to users, the rate achieved at secondary users and the interference at the primary nodes. The proposed framework takes into account the imperfect knowledge of the channels between the secondary and the primary nodes and also mitigates the interference caused by the primary users at secondary receivers. In order to solve the underlying NP-hard problem, we propose a three-step algorithm with two alternative schemes for power allocation, namely: equal power and equal rate. In addition, we provide a solution by reducing the problem to an Integer Linear Program (ILP). The performances of equal rate, equal power, and ILP methods are studied in under different constraints. 

\end{abstract}
\IEEEpeerreviewmaketitle


\section{Introduction}
\label{sec:Introduction}
In cognitive radio (CR) networks, unlicensed secondary users (SUs) co-exist and share the wireless spectrum with licensed primary users (PUs) without causing significant interference to the PUs. There are three paradigms of co-existence of SUs and PUs in a CR network: interweave, overlay and underlay \cite{Goldsmith2012}. In this paper, we consider an underlay CR network, where the secondary transmitter is allowed to transmit as long as the interference temperature at primary users (PUs) is below a certain threshold. The interference temperature at the PUs can be controlled by using multiple antennas at the SU transmitter. The multiple-input multiple-output (MIMO) technique allows SU transmitter to employ beamforming to control the interference power transmitted towards the PUs in the network. Further, in multi-user MIMO scenario, by using a large number of antennas at the SU transmitter (massive MIMO), a high number of SU receivers can be served. Various recent research works, such as \cite{Noam2012,Tsinos2013,Xu2013,Du2012,Du2013}, address the topic of MIMO CR in underlay scenario. However, in \cite{Noam2012}, a feedback mechanism is required from PU to SU in order to estimate the interference, while \cite{Xu2013} requires the perfect knowledge of the channel between the SU transmitter and the PU in order to control the interference. Transmission to multiple SU receivers is not considered in \cite{Tsinos2013}. Papers \cite{Du2012,Du2013} do not consider interference caused by the PU transmitter at the SU receiver.

In this paper, we consider an underlay cognitive cellular network where a secondary base-station (SU BS) and multiple secondary users (SU UEs) coexist along with PUs. We propose an optimization framework to maximize the number of SU UEs in downlink transmission from the SU BS, which employs a large number of transmit antennas. The constraints of the optimization are as follows: 1) the total power allocated to all SU UEs is limited to $P^0$, 2) the minimum rate achieved at any SU UE selected for downlink transmission is $R^0$, and 3) the interference power at PUs cannot exceed $I^0$. The proposed framework takes into account the imperfect knowledge of the channel between SU BS and PU in line-of-sight (LOS) environment. Further, the optimization framework, unlike \cite{Du2012}, and \cite{Du2013}, also includes receiver beamforming at SU UEs in order to mitigate the interference caused by PU transmitter. 

This paper is organized as follows. The system model is presented in Section \ref{sec:Model}. The error model for the channel estimation is presented in Section \ref{sec:channel}, while the optimization problem is presented in Section \ref{sec:optimization}. The proposed optimization problem  is a mixed integer program and a NP-hard problem. We propose a three-step algorithm to solve the transceiver beamforming and admission control problem in Section \ref{sec:algorithm}. In this section, we also provide a solution to the optimization problem by reducing it to an Integer Linear Program (ILP). The simulation results are presented in Section \ref{sec:Results}. Finally, Section \ref{sec:conclusion} concludes this paper.

\section{System Model}
\label{sec:Model}
\begin{figure}
\centering
\includegraphics[width=0.7 \columnwidth]{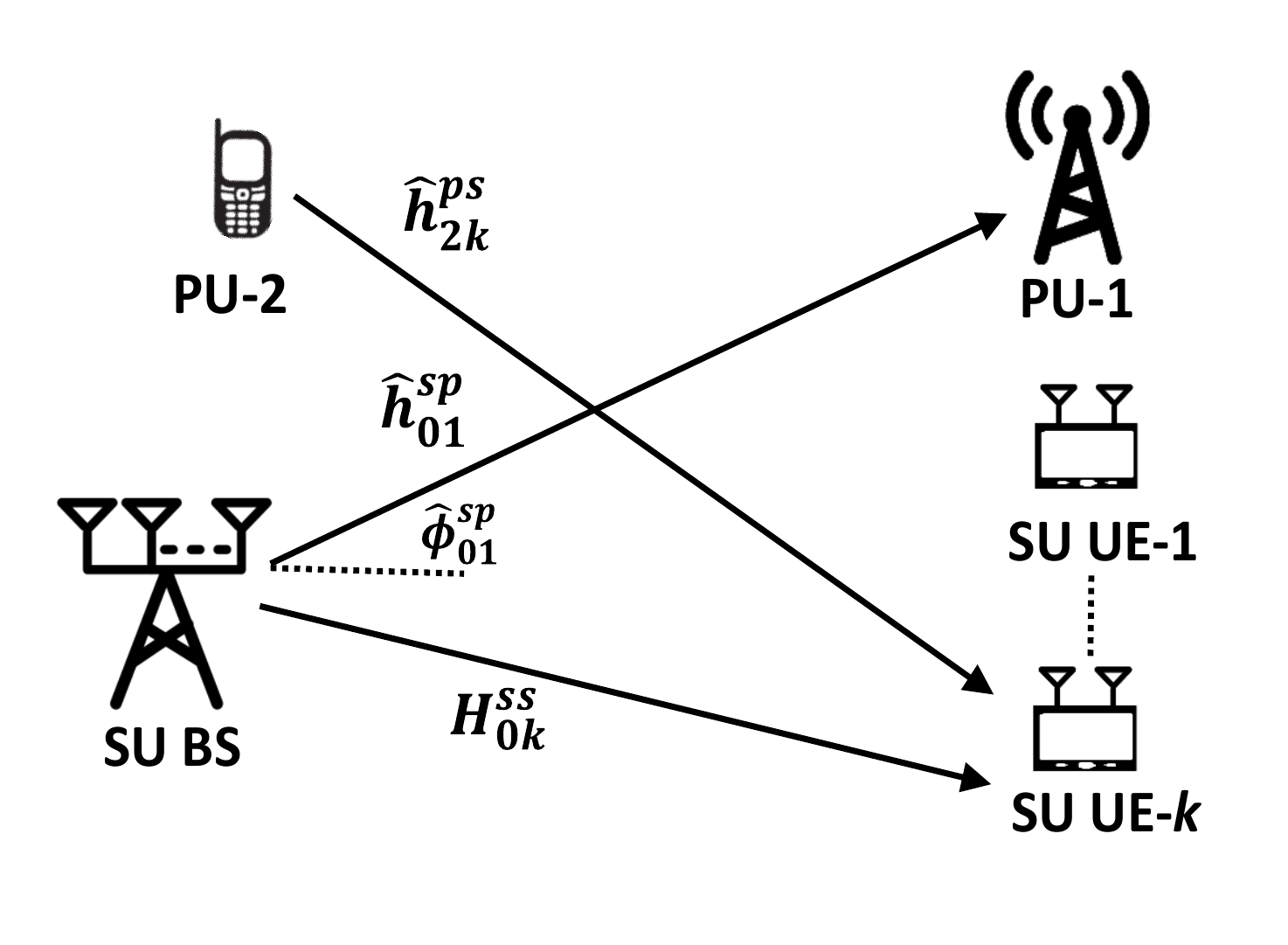}
\caption{System Model: Massive MIMO Cognitive Radio Network.}
\label{fig:system_model}
\vspace{-5mm}
\end{figure}

As shown in Fig. \ref{fig:system_model}, we consider a CR network with one SU BS and $K$ SU UEs denoted by SU UE-$k$, $k=1,2,...,K$. The number of antennas at SU BS and SU UEs are denoted by $M_b$ and $M_u$, respectively,  with $M_b \gg K$. There are two primary nodes in the network, denoted by PU-1 and PU-2. Both PU-1 and PU-2 have one antenna each. The primary nodes switch the roles from transmitter to receiver and vice-versa. 

We consider reciprocal, and line-of-sight (LOS) channels between any two nodes in the network.  Let $\theta^{ss}_{0k}$ be the angle at which SU UE-$k$ is located with respect to SU BS and let the corresponding angular sine be $\phi^{ss}_{0k} = \sin(\theta^{ss}_{0k})$. Then, the MIMO channel between SU BS and SU UE-$k$ is $H^{ss}_{0k} = \beta^{ss}_{0k}  a_{M_{u}}(\phi^{ss}_{0k})a_{M_{b}}^H(\phi^{ss}_{0k}) $, where $\beta^{ss}_{0k} = \alpha^{ss}_{0k}e^{j\psi^{ss}_{0k}}$ and $\alpha^{ss}_{0k}$, and  $\psi^{ss}_{0k}$ are the channel attenuation and phase, respectively \cite{Ngo2014}. Let the steering vectors at SU BS and SU UE-$k$ be $a_{M_{b}}(\phi) = [1, e^{j\pi \phi}, e^{j\pi 2 \phi},...e^{j\pi (M_{b}-1)\phi}]^H$ and $a_{M_{u}}(\phi)=[1, e^{j\pi \phi}, e^{j\pi 2 \phi},...e^{j\pi (M_{u}-1)\phi}]^H$, respectively. We assume that the channel attenuation $\alpha^{ss}_{0k}$ depends on the path-loss, while the phase $\psi^{ss}_{0k}$ is a uniform random variable in $[0, 2\pi]$. Similarly, MISO channel between SU BS and PU-2 is $h^{sp}_{02} = \beta^{sp}_{02} a_{M_{b}}^H(\phi^{sp}_{02})$, where $ \beta^{sp}_{02} = \alpha^{sp}_{02} e^{j\psi^{sp}_{02}}$, and SIMO channel between PU-1 and SU UE-$k$ is $h^{ps}_{1k} = \beta^{ps}_{1k} a_{M_{u}}(\phi^{ps}_{1k})$, where $\beta^{ps}_{1k} = \alpha^{ps}_{1k}e^{j\psi^{ps}_{1k}}$.

\section{Channel Estimation}
\label{sec:channel}
We assume that SU BS and SU UEs have the knowledge of the channel $H^{ss}_{0k}$. However, SU BS does not have a priori knowledge of channels to PUs: $h^{sp}_{01}$ and $h^{sp}_{02}$. In order to estimate and control the interference power at PU-1 and PU-2, SU BS needs to estimate channel attenuations $\alpha^{sp}_{01}$ and $\alpha^{sp}_{02}$ and angular sines $\phi^{sp}_{01}$ and $\phi^{sp}_{02}$. 

We assume that the channel attenuation is computed by finding the location of PU-1 using a non-cooperative localization algorithm, such as Weighted Centroid Localization (WCL) \cite{Wang2011}. The PU-1 location estimates are then used to obtain the distance $\hat{d}^{sp}_{01}$ between SU BS and PU-1. The channel attenuation is computed using the path-loss formula: $\hat{\alpha}^{sp}_{01}=(\hat{d}^{sp}_{01})^{-\gamma/2}$, where $\gamma$ is the path-loss exponent. Similarly, $\hat{\alpha}^{sp}_{02}$ is obtained. The error in the estimation of channel attenuations is modeled as a zero mean Gaussian random variable with variance $\sigma_\alpha^2$.

Further, in order to estimate the angular sine, $\phi^{sp}_{01}$, we assume that SU BS scans the angular space $\phi \in [-1,1]$ at $M_b$ equally spaced grid angles when PU-1 is transmitting. The estimated angle $\hat{\phi}^{sp}_{01}$ is the grid angle at which maximum energy is received. Since the grid angles are separated by $2/M_b$, the error in estimation of angular sines is uniformly distributed in range $[-1/M_b,1/M_b]$. Channel attenuation $\hat{\alpha}^{sp}_{02}$ and angular sine $\hat{\phi}^{sp}_{02}$ are estimated in similar way when PU-2 is transmitting the signal.

\section{Optimization Framework}
\label{sec:optimization}

The goal of the proposed optimization framework is to find transmitter and receiver beamforming vectors $v_k$ and $u_k$, respectively and power allocation $P_k$ in order to maximize the number of SU UEs in downlink transmission with given minimum rate requirement, $R^0$, while keeping the interference to PUs below a threshold $I^0$ with maximum available power constraints $P^0$. The optimization problem is presented in (\ref{eq:optimization_problem_start})-(\ref{eq:optimization_problem_end}).

In (\ref{eq:sk}), the variable $s_k \in \{0,1\}$ indicates whether SU UE-$k$ is selected for transmission ($s_k=1$) or not ($s_k=0$). All the SU UEs included in downlink transmission have the minimum channel capacity $R^0$ as indicated in (\ref{eq:rate_constraints}), where $\sigma_w^2$ is the noise power. The transmitter beamforming constraints in (\ref{eq:tx_beamforming_1}) ensure that the beamforming vector $v_k$ corresponding to SU UE-$k$ has nulls in the directions of other SU UEs ($s_l{H}^{ss}_{0l}v_k =0,  k \neq l$). (Note: The constraint $||v_k||^2 = s_k$ ensures that $v_k = 0$ for $s_k=0$). Further, the beamforming vector $v_k$ also has nulls in the estimated directions of PU-1 and PU-2 ($\hat{h}^{sp}_{01}v_k =0, \text{ } \hat{h}^{sp}_{02}v_k =0$). However, due to the angle estimation error $\hat\phi^{sp}_{01}-\phi^{sp}_{01}$, the constraints in (\ref{eq:tx_beamforming_1}) do not place a null at the actual angle $\phi^{sp}_{01}$. Therefore, the SU BS needs to estimate the interference power at PU-1 in order to keep it below $I^0$. The average interference power at PU-1 is estimated as follows. The power transmitted by SU BS in any arbitrary direction $\phi$ is $\sum_{k = 1}^{K} P_k v_k^H a_{M_b}({\phi})a_{M_b}^H({\phi})v_k$. Since the estimation error $\phi^{sp}_{01} - \hat{\phi}^{sp}_{01}$ is uniformly distributed in $[- \frac{1}{M_b}, \frac{1}{M_b}]$, the true value of the angle $\phi^{sp}_{01}$ lies in the interval $[\hat{\phi}^{sp}_{01} - 1/M_b, \hat{\phi}^{sp}_{01} + 1/M_b]$ with uniform probability $2/M_b$. The average interference power transmitted in the interval $[\hat{\phi}^{sp}_{01} - 1/M_b, \hat{\phi}^{sp}_{01} + 1/M_b]$ is $\sum_{k = 1}^{K} P_k v_k^H F(\hat{\phi}^{sp}_{01})v_k$, where
\begin{equation}
F(\hat{\phi}^{sp}_{01}) = \frac{M_b}{2}\int_{\hat{\phi}^{sp}_{01} - 1/M_b}^{\hat{\phi}^{sp}_{01} + 1/M_b} a_{M_b}(\phi)a_{M_b}^H(\phi) d\phi. 
\label{eq:F}
\end{equation}
Therefore, the estimated interference power at PU-1 is $(\hat{\alpha}^{sp}_{01})^2 \sum_{k =1}^{K} P_k v_k^H F(\hat{\phi}^{sp}_{01})v_k$. Similar expression is obtained for the estimated interference power at PU-2. The interference constraints corresponding to PU-1 and PU-2 are included in constraints (\ref{eq:pu1_int}) and (\ref{eq:pu2_int}). 
\begin{align}
&\max_{\{u_k,v_k,P_k, s_k\}}  \sum_{k=1}^K s_k, 
\label{eq:optimization_problem_start}
\\&\text{subject to:} ~~ s_k   \in \{0,1\}, \label{eq:sk}
\\ & R_k= \log_2\left(1 + \frac{P_k |u_k^H H^{ss}_{0k}v_k|^2}{\sigma_w^2}\right)\geq s_k R^0,
\label{eq:rate_constraints}
\\ & \hat{h}^{sp}_{01}v_k =0, \text{ } \hat{h}^{sp}_{02}v_k =0,\text{ }||v_k||^2 = s_k, \text{ }  s_l{H}^{ss}_{0l}v_k =0,  k \neq l,
\label{eq:tx_beamforming_1}
\\ & (\hat{\alpha}^{sp}_{01})^2 \sum_{k =1}^{K} P_k v_k^H F(\hat{\phi}^{sp}_{01})v_k \leq I^0,\label{eq:pu1_int}
\\ & (\hat{\alpha}^{sp}_{02})^2 \sum_{k =1}^{K} P_k v_k^H F(\hat{\phi}^{sp}_{02})v_k \leq I^0, \label{eq:pu2_int}
\\&\sum_{k =1}^K s_k P_k \leq P^0,  \text{ } P_k \geq 0,
\\ & u_k^H \hat{h}^{ps}_{1k} = 0,\text{ } u_k^H \hat{h}^{ps}_{2k} = 0, \text{ } ||u_k||^2 = s_k.
\label{eq:optimization_problem_end}
\end{align}

\section{Algorithm}
\label{sec:algorithm}
The optimization problem described in (\ref{eq:optimization_problem_start})-(\ref{eq:optimization_problem_end}) is a mixed integer program and therefore, a NP-hard problem. We propose a three-step algorithm to solve the problem. The three steps are: 1) transmitter beamforming 2) receiver beamforming 3) power allocation and interference control. It should be noted that the term $\Gamma_k=|u_k^H H^{ss}_{0k}v_k|^2$ in (\ref{eq:rate_constraints}) represents the equivalent channel strength between SU BS and SU UE-$k$. Since our objective is to serve maximum number of SU UEs with given rate constraints, we maximize the term $\Gamma_k$ for $k={1,2,...,K}$ with constraints (\ref{eq:tx_beamforming_1}) and (\ref{eq:optimization_problem_end}) in transmitter and receiver beamforming steps, respectively. In step 1 and 2, it is assumed that all the SU UEs are selected for transmission i.e. $s_k=1, \forall k$.
 The problem is optimized for variables $P_k$ and $s_k$ in step 3. The algorithmic steps are described below.
\subsubsection{Transmitter Beamforming}
\label{sec:transmitter_beamforming}
The term $\Gamma_k$ can be decomposed using $H^{ss}_{0k} = \alpha^{ss}_{0k} e^{j\psi^{ss}_{0k}} a_{M_{u}}(\phi^{ss}_{0k})a_{M_{b}}^H(\phi^{ss}_{0k})$, as:
\begin{equation}
\nonumber \Gamma_k = |u_k^H H^{ss}_{0k}v_k|^2 = (\alpha^{ss}_{0k})^2 |u_k^H a_{M_u}({\phi}^{ss}_{0k})|^2 |a_{M_b}^H({\phi}^{ss}_{0k}) v_k|^2 
\end{equation}

In transmitter beamforming, $|a_{M_b}^H({\phi}^{ss}_{0k}) v_k|^2$ is maximized subject to constraints (\ref{eq:tx_beamforming_1}). The solution is provided by nullsteering-beamforming algorithm \cite{Friedlander1989}, as shown in  Algorithm \ref{algorithm}. 
\subsubsection{Receiver Beamforming}
\label{sec:receiver_beamforming}
In receiver beamforming, $|u_k^H a_{M_u}({\phi}^{ss}_{0k})|^2$ is maximized subject to constraints (\ref{eq:optimization_problem_end}). The solution is provided by nullsteering-beamforming algorithm, as shown in  Algorithm \ref{algorithm}. In steps 1 and 2, we assume that the channels matrices $H_{0k}, k=1,2,...K$ are linearly independent of each other and are also linearly independent of  $\hat{h}^{sp}_{01}$, and $\hat{h}^{sp}_{02}$. This means that aligning the steering vector $v_k$ to SU UE-$k$ while placing the nulls towards all the other SU UEs and PUs is not infeasible. Therefore, setting $s_k=1, \forall k$ in oder to maximize $\Gamma_k = |u_k^H H^{ss}_{0k}v_k|^2$ with constraints (\ref{eq:tx_beamforming_1}) and (\ref{eq:optimization_problem_end}), does not result in an infeasible problem. The required condition of channel independence holds in a LOS environment if the angles of SU UEs ($\phi_{0k}^{ss}$) and PUs ($\hat{\phi}^{sp}_{01}$ and $\hat{\phi}^{sp}_{02}$) are different from each other. 

\begin{algorithm}[t]
	\caption{}
	\label{algorithm}
	\begin{algorithmic}[1]
		\State Receiver Beamforming:  $\forall k$, set $s_k=1$ 
		\Statex Maximize $|u_k^H a_{M_u}(\phi^{ss}_{0k})|^2$ s. t. constraints (\ref{eq:optimization_problem_end}).
		\Statex Solution: $u_k=\frac{(I-P_A)a_{M_u}(\phi^{ss}_{0k})}{||(I-P_A)a_{M_u}(\phi^{ss}_{0k})||}$, where $P_A=A(A^HA)^{-1}A^H$  and $A=[a_{M_u}(\hat{\phi}^{ps}_{1k}), a_{M_u}(\hat{\phi}^{ps}_{2k})]$.
		\State Transmitter Beamforming: $\forall k, l = [1,2...K], l \neq k$, 
		\Statex Maximize $|a_{M_b}^H(\phi^{ss}_{0k}) v_k|^2$ s. t. 
		constraints (\ref{eq:tx_beamforming_1}).
		\Statex Solution: $v_k=\frac{(I-P_B)a_{M_b}(\phi^{ss}_{0k})}{||(I-P_B)a_{M_b}(\phi^{ss}_{0k})||}, P_B=B(B^HB)^{-1}B^H$ and  $B=[a_{M_b}(\hat{\phi}^{sp}_{01}), a_{M_b}(\hat{\phi}^{sp}_{02})...a_{M_b}({\phi}^{ss}_{0l})..]$
		\State Power Allocation: (use either scheme 1 or scheme 2)
		\Statex Scheme 1: Equal Power
		\begin{algorithmic}
			\While{$\sum_{k=1}^{K} s_k \geq 1$}
			\State $P_k \gets P^0/\sum_{j=1}^{K} s_j$
			\If{$R_k < s_k R^0$ for any $k$}
			\State $s_j = 0, v_j = 0, u_j = 0$, where $ j = {\text{arg}}\mathop{\min}\limits_{k} R_k$.
			\EndIf
			\EndWhile
		\end{algorithmic}
		\Statex Scheme 2: Equal Rate
		\begin{algorithmic}
			\State $P_k \gets \frac{\sigma_w^2 (2^{R^0}-1)}{|u_k^H H^{ss}_{0k}v_k|^2}$
			\While{$\sum_{k=1}^{K} s_k \geq 1$}
			\If{$\sum_{k=1}^{K} s_kP_k > P^0$}
			\State  $s_j = 0, v_j = 0, u_j = 0$, where $j = \text{arg}\mathop{\max}\limits_{k} P_k$.
			\EndIf
			\EndWhile
		\end{algorithmic}
	\end{algorithmic}
	\begin{algorithmic}[1]
		\Statex Interference Control:
		\begin{algorithmic}
			\If{Constraint (\ref{eq:pu1_int}) is not satisfied}\\
			$s_j = 0, v_j = 0, u_j = 0$, where $j =  \text{arg}\mathop{\min}\limits_{k} |\phi^{ss}_{0k} - \hat{\phi}^{sp}_{01}|$.
			\EndIf
			\If{Constraint (\ref{eq:pu2_int}) is not satisfied} \\
			$s_j = 0, v_j = 0, u_j = 0$, where $j =  \text{arg}\mathop{\min}\limits_{k} |\phi^{ss}_{0k} - \hat{\phi}^{sp}_{02}|$.
			\EndIf
		\end{algorithmic}
	\end{algorithmic}
\end{algorithm}

\subsubsection{Power Allocation and Interference Control}
\label{sec:power_allocation}
We propose two schemes for power allocation: equal power and equal rate. In addition, we also provide a solution by reducing the problem in an Integer Linear Program (ILP). The transmit- and receive-beamforming vectors, $u_k$ and $v_k$, obtained in the above steps are treated as constants in this step.

In the equal power scheme, the total available power is distributed equally among the selected SU UEs. Further, if the rate constraints are not satisfied with given power allocation, the SU UE which achieves the minimum rate is removed (by setting $s_k=0$) from the downlink transmission. This is because the SU UE achieving the minimum rate despite maximizing $\Gamma_k$ in previous steps, has a weak channel $H_k$ and will require high power to achieve the rate $R^0$. Further, the additional power made available by dropping this SU UE is equally redistributed among the remaining SU UEs and the rate constraints are checked again. The scheme is presented in Algorithm \ref{algorithm}.

In the equal rate scheme, the power is allocated such that each SU UE achieves equal rate of $R^0$. Further, if the total power constraints ($P^0$) are not satisfied, the SU UE that is allocated the maximum power is removed  (by setting $s_k=0$) from downlink transmission. Again, the reason for dropping this particular SU UE is that it has a weak channel $H_k$ which results in high power consumption. Finally, the power constraints are checked again and if they are satisfied, the interference control step is implemented.

In the interference control step, the interference constraints in (\ref{eq:pu1_int}) and (\ref{eq:pu2_int}) corresponding to PU-1 and PU-2 are tested. If the constraint is not satisfied, the SU UE whose angular sine is closest to the angular sine of PU-1 (i.e.$|\phi^{ss}_{0k} - \hat{\phi}^{sp}_{01}|$ is minimum) or PU-2, is removed from downlink transmission. This is because the downlink signal transmitted to such SU UE causes the maximum interference at PUs due to proximity in angular domain.

\paragraph{Power Allocation and Interference Control using ILP}
\label{sec:ilp}
In step 3 of the algorithm, by considering $v_k$ and $u_k$ as constants, the optimization problem (\ref{eq:optimization_problem_start})-(\ref{eq:optimization_problem_end}) can be reformulated as shown in (\ref{eq:optimization_problem_start2})-(\ref{eq:optimization_problem_end2}). 
\begin{align}
\max_{\{P_k, s_k\}}  &\sum_{k=1}^K s_k, 
\label{eq:optimization_problem_start2}
\\\text{subject to:}~ & s_k   \in \{0,1\},
\label{eq:sk2}
\\ & R_k= \log_2\left(1 + \frac{P_k \Gamma_k}{\sigma_w^2}\right)\geq R^0,
\label{eq:rate_constraints2}
\\ & (\hat{\alpha}^{sp}_{0j})^2 \sum_{k =1}^{K} s_kP_k g_{k,j}  \leq I^0, j=1,2,
\label{eq:pu1_int2}
\\& \sum_{k =1}^K s_k P_k \leq P^0,  \text{ } P_k \geq 0
\label{eq:optimization_problem_end2}
\end{align}
In the problem, $g_{k,1} = v_k^H F(\hat{\phi}^{sp}_{01})v_k$ and $g_{k,2} = v_k^H F(\hat{\phi}^{sp}_{02})v_k$. It should be noted that $g_{k,1}$ and $g_{k,2}$ are non-negative because $F(\hat{\phi}^{sp}_{01})$ and $F(\hat{\phi}^{sp}_{02})$ are positive semi-definite matrices. From (\ref{eq:rate_constraints2}), it is clear that the minimum feasible power allocation for the $k^{th}$ user in order to achieve the rate $R^0$ is $P_k= \sigma_w^2(2^{R^0}-1)/\Gamma_k$. Further, we note that $(\hat{\alpha}^{sp}_{01})^2$, $s_k$, $g_{k,1}$, $(\hat{\alpha}^{sp}_{02})^2$, and $g_{k,2}$ are non-negative. Therefore, if there exists any feasible power allocation that satisfies constraints (\ref{eq:rate_constraints2})-(\ref{eq:optimization_problem_end2}) for given $s_k's$, then $P_k= \sigma_w^2(2^{R^0}-1)/\Gamma_k$ is also a feasible power allocation satisfying constraints (\ref{eq:rate_constraints2})-(\ref{eq:optimization_problem_end2}). Further, since the objective function is independent of $P_k$, $P_k$'s can be considered as constants with $P_k= \sigma_w^2(2^{R^0}-1)/\Gamma_k$ and the problem is reduced to an Integer Linear Program (ILP) with binary variables $s_k, k={1,2,...,K}$, as shown in (\ref{eq:optimization_problem_start3})-(\ref{eq:optimization_problem_end3}). This problem can now be solved using standard ILP solvers.
\begin{align}
\max_{\{s_k\}}  & \sum_{k=1}^K s_k, 
\label{eq:optimization_problem_start3}
\\ \text{subject to:} ~ & s_k   \in \{0,1\},
\label{eq:sk3}
\\ & (\hat{\alpha}^{sp}_{0j})^2 \sum_{k =1}^{K} s_kP_k g_{k,j}  \leq I^0, j=1,2,
\label{eq:pu1_int3}
\\ & \sum_{k =1}^K s_k P_k \leq P^0,  \text{ } P_k \geq 0
\label{eq:optimization_problem_end3}
\end{align}

Finally, let us consider a case where a feasible power allocation that satisfies constraints (\ref{eq:rate_constraints2})-(\ref{eq:optimization_problem_end2}) does not exist. In this case also, we set $P_k= \sigma_w^2(2^{R^0}-1)/\Gamma_k$ and solve the ILP (\ref{eq:optimization_problem_start3}) -(\ref{eq:optimization_problem_end3}). The solution of the ILP, in this case, is $s_k = 0, \forall k$.


\section{Simulation Results}
\label{sec:Results}
We consider a CR network spanning over a square-shaped area of $100$m $\times$ $100$m with SU BS located at the origin with x- and y-coordinates $[0, 0]$. The location coordinates of PU-1 and PU-2 are $[20,20]$ and $[-30, 30]$ respectively. The proposed algorithm is simulated for 1000 iterations, with $K$ SU UEs uniformly distributed in the square-shaped area. The variance of error $(\sigma_\alpha^2)$ in estimation of channel coefficients, $\alpha$, is assumed to be  1\% of the true value of $\alpha$.

\subsection{Impacts of total number of SU UEs and number of antennas}
\label{sec:impact_K}
\begin{figure}[t!]
	\centering
	\includegraphics[width=0.6 \columnwidth]{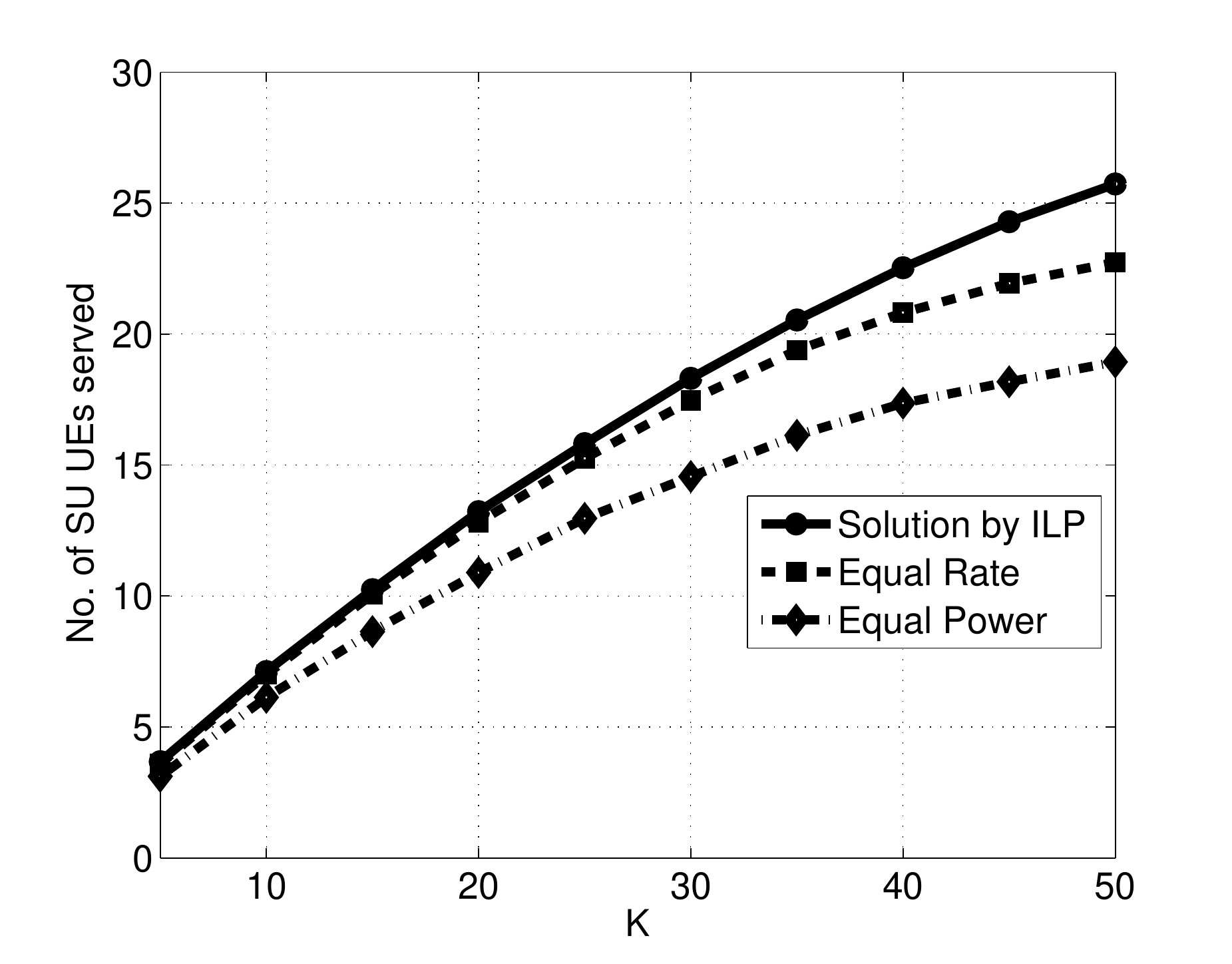}
	\caption{Number of SU UEs served vs. $K$. Parameters: $M_b = 128, M_u = 4$. $I^0 = 0 $ dBm, $P^0 = 60 $ dBm, $R^0 = 1$ bps/Hz, $\sigma_w^2=0$ dBm.}
	\label{fig:no_sus_served_K}
	\vspace{-5mm}
\end{figure}

The number of SU UEs selected is plotted against the total number of SU UEs in the network in Fig. \ref{fig:no_sus_served_K}. It can be observed that the performance of ILP, equal rate, and equal power schemes is comparable if number of users in the network is small ($K<10$). As the number of users in the network increases, the equal power serves fewer users as compared to equal rate and ILP. The equal rate serves more SU UEs than equal rate for the following reason. In both equal rate and equal power schemes, the SU UEs are dropped (by setting $s_k=0$) in power allocation and interference control steps. The interference control procedure is the same for both schemes. Therefore, let us consider the power allocation step in the two schemes. In equal rate scheme, the SU UE that is allocated the maximum power is dropped in the power allocation step. Therefore, the cause of interference to PU (SU UE with maximum power) has been removed in power allocation step itself and the interference control step may not be required. On the other hand, in the power allocation step of equal power scheme, the SU UE that achieves the minimum rate is dropped. It should be noted that the SU UE achieving the minimum rate may not cause significant interference to PUs, therefore additional SU UEs are dropped in interference control step in equal power scheme, which results in dropping additional SU UEs from the downlink transmission.

\begin{figure}[t!]
	\centering
	\includegraphics[width=0.6 \columnwidth]{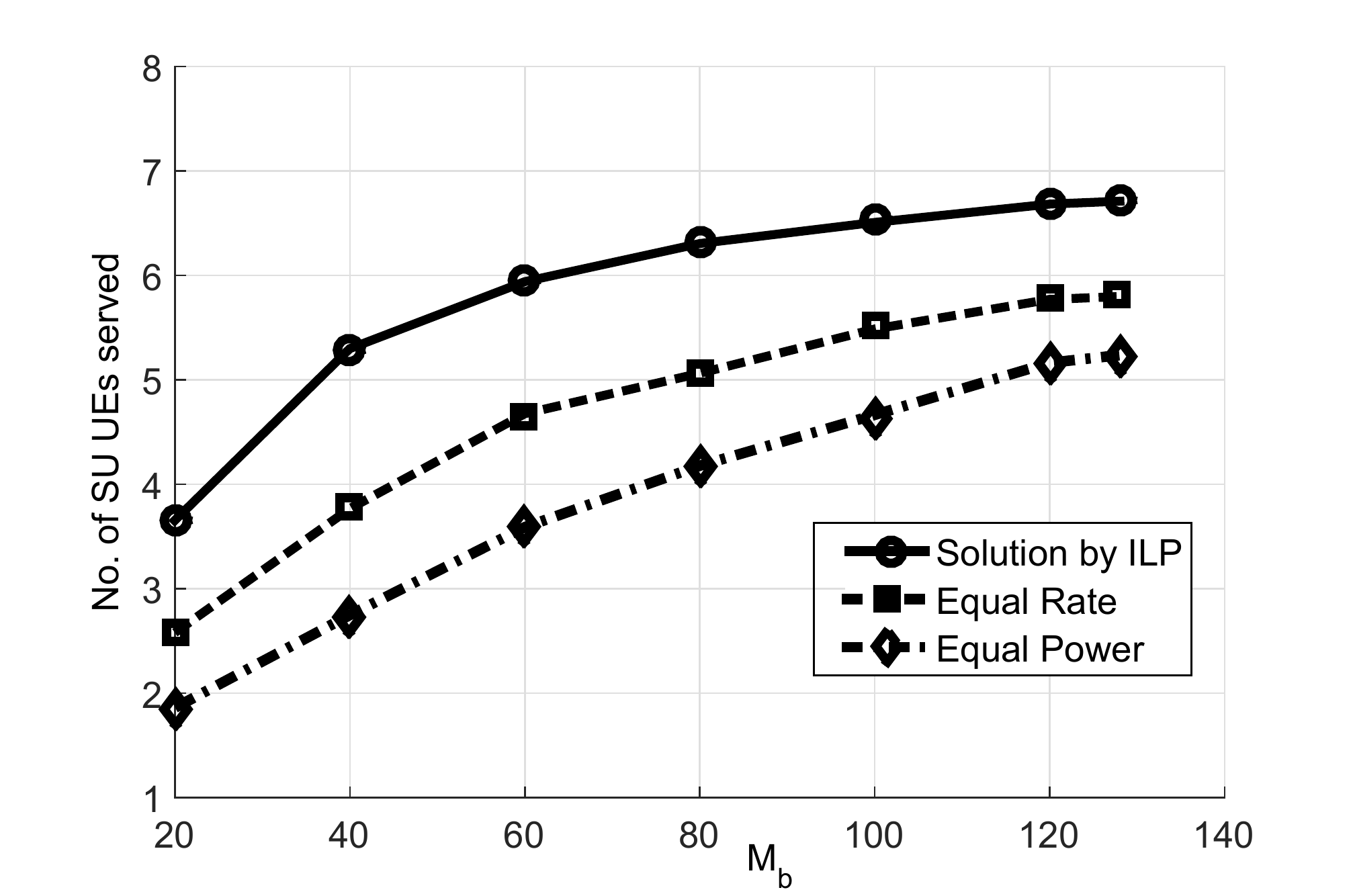}
	\caption{Number of SU UEs served vs. $M_b$. Parameters: $K = 10, M_u = 4$. $P^0 = 60 $ dBm,  $I^0 = 0$ dBm,  $\sigma_w^2=0$ dBm, $R^0 = 1$ bps/Hz.}
	\label{fig:no_sus_served_Mb}
	\vspace{-5mm}
\end{figure}

It can be observed that the number of SU UEs increase with higher number of antennas at SU BS in Fig. \ref{fig:no_sus_served_Mb}. Increasing $M_b$ not only provides more transmission opportunities in angular domain, but also reduces the error in PU angle estimation, which results in higher SU UEs in downlink transmission.

\subsection{Impact of the constraints $P^0$, $I^0$, $R^0$}
\label{sec:impact_constraints}
\begin{figure}[t!]
	\centering
	\includegraphics[width=0.6 \columnwidth]{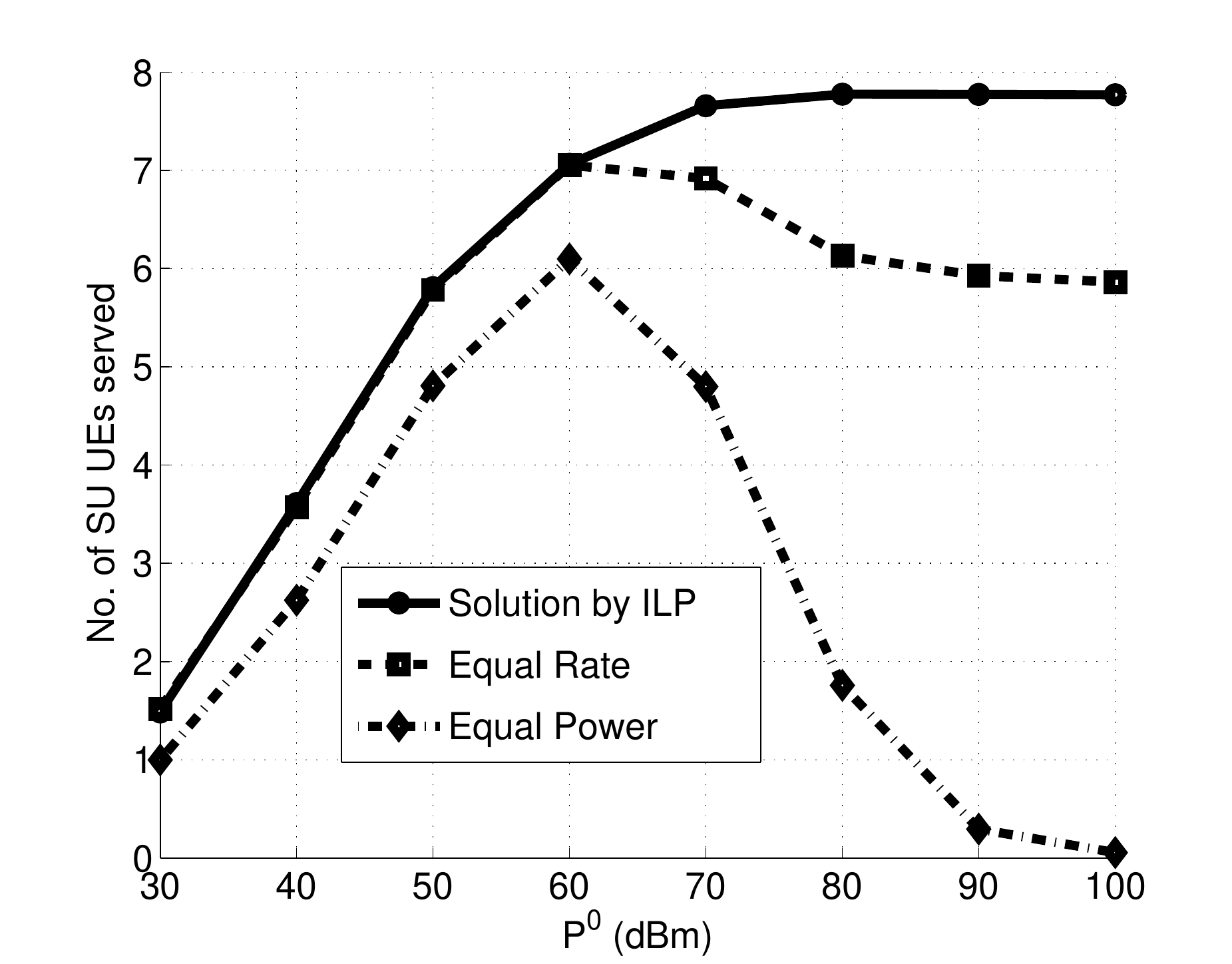}
	\caption{Number of SU UEs served vs. $P^0$. Parameters: $K = 10, M_b = 128, M_u = 4$. $I^0 = 0 $ dBm,  $R^0 = 1$ bps/Hz, $\sigma_w^2=0$ dBm.}
	\label{fig:no_sus_served_P0}
	\vspace{-5mm}
\end{figure}

In order to study the impact of the constraints on the number of SU UEs supported in the downlink, the total number of SU UEs in the network in fixed to $K=10$. The impact of increasing the total power constraint $P^0$ is shown in Fig. \ref{fig:no_sus_served_P0}. It is observed that the solution provided by ILP matches with equal rate scheme for $P^0\leq 60$ dBm. Equal power scheme served the least number of users among the three schemes. Especially as $P^0$ approaches $100$ dBm, the number of users served by equal power scheme reduces to zero. At higher $P^0$, the power allocated to each user in equal power scheme is also high. For example, $P_0=100$ dBm results in $P_k=90$ dBm for each user.  This results in a significant interference to the PUs. Therefore, most of the users are dropped from downlink transmission in the interference control step.

\begin{figure}[t!]
	\centering
	\includegraphics[width=0.6 \columnwidth]{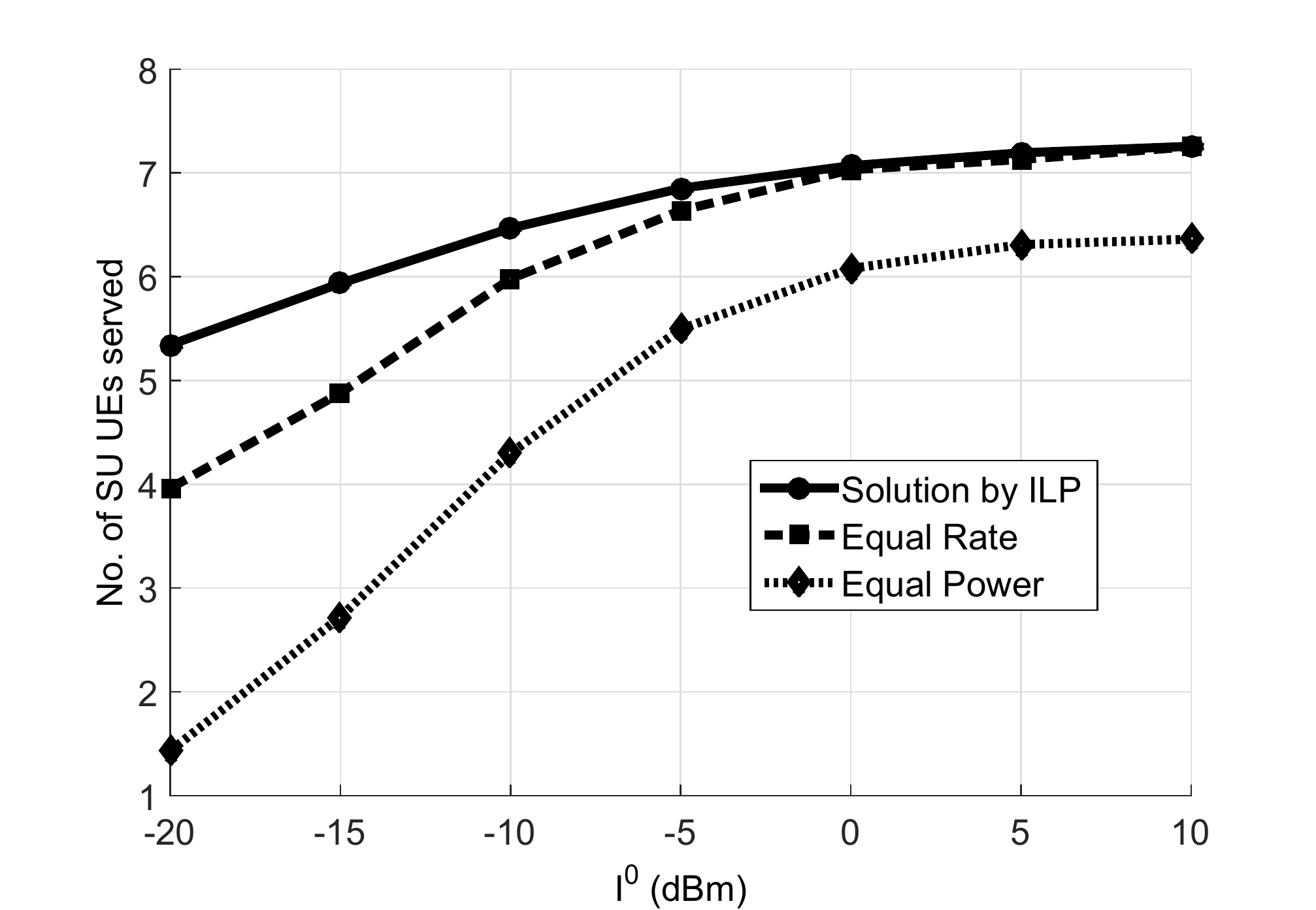}
	\caption{Number of SU UEs served vs. $I^0$. Parameters: $K = 10, M_b = 128, M_u = 4$. $P^0 = 60 $ dBm,  $R^0 = 1$ bps/Hz, $\sigma_w^2=0$ dBm.}
	\label{fig:no_sus_served_I0}
	\vspace{-5mm}
\end{figure}

\begin{figure}[t!]
	\centering
	\includegraphics[width=0.6 \columnwidth]{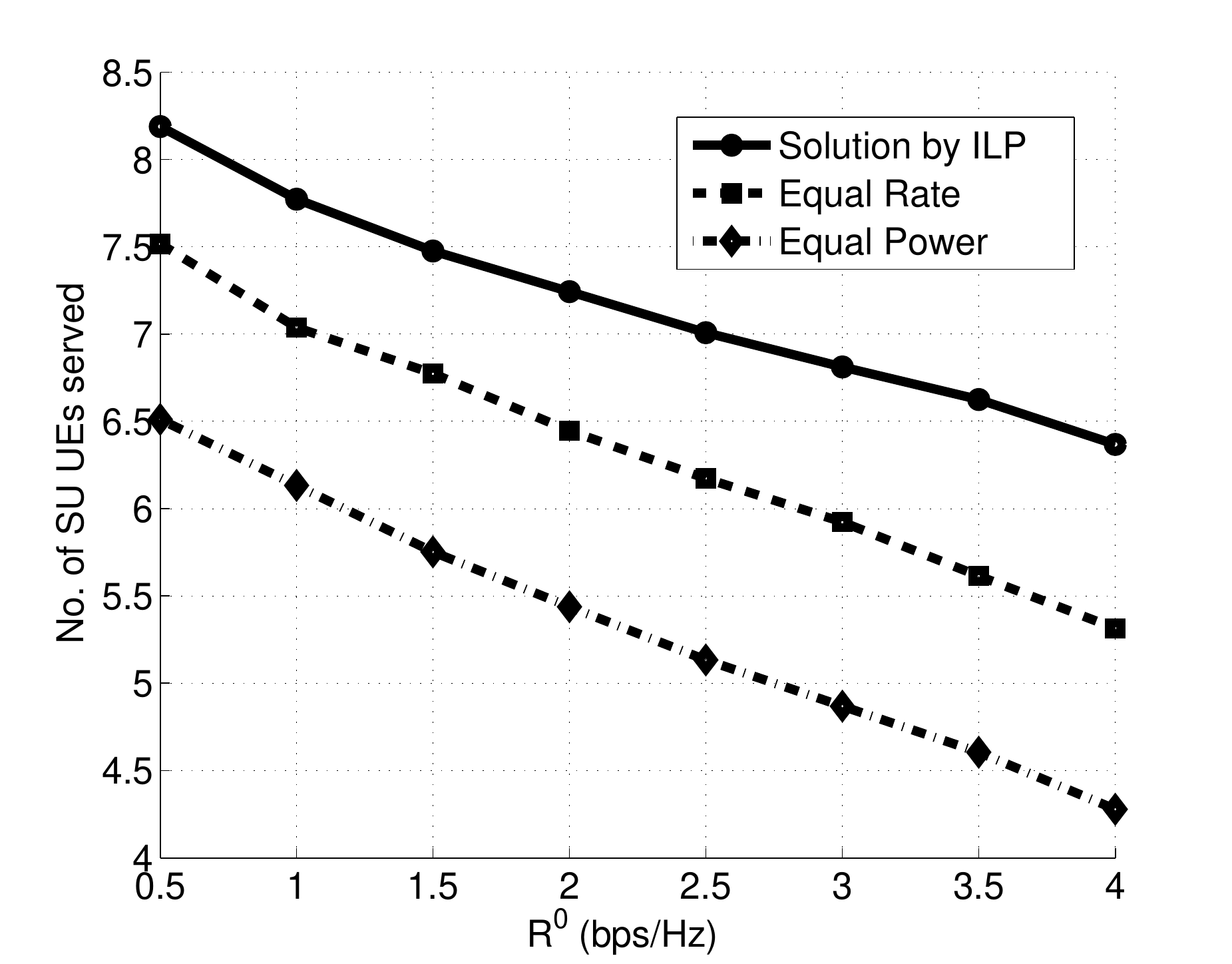}
	\caption{Number of SU UEs served vs. $R^0$. Parameters: $K = 10, M_b = 128, M_u = 4$. $P^0 = 60 $ dBm,  $I^0 = 0$ dBm,  $\sigma_w^2=0$ dBm.}
	\label{fig:no_sus_served_R0}
	\vspace{-5mm}
\end{figure}

The impact of increasing interference constraint $I^0$ is shown in Fig. \ref{fig:no_sus_served_I0}. It can be observed that the performance of equal rate scheme matches that of ILP for $I^0\geq 0$ dBm. However, in a stricter interference requirement ($I^0=-20$ dBm), the interference control step becomes more important in terms of selecting the SU UEs for downlink transmission. Therefore, the optimum solution provided by ILP is significantly better in terms of serving more user as compared to equal rate scheme. Finally, as shown in Fig. \ref{fig:no_sus_served_R0}, the number of users served in ILP, equal rate, and equal power reduces linearly with increased rate requirement, as expected due to limited available  power. 

\section{Conclusion and Future Work}
\label{sec:conclusion}
In this paper, we have proposed an optimization framework to maximize the number of SU UEs in downlink transmission from SU BS in a CR network. The optimization problem is constrained by the total power allocated to SU UEs, interference power received at PUs, and minimum rate achieved by the SU UEs selected for the transmission. A three-step algorithm has been proposed to solve the optimization problem using two alternative power allocation schemes, namely equal rate and equal power. In addition, we have also provided a solution by reducing the problem to an ILP. It has been observed that the performance of equal rate and equal power schemes is comparable if the total number of SU UEs is small. For all other scenarios studied, the equal rate scheme serves more SU UEs than equal power. It has been observed that the performance of equal rate matches with the solution provided by ILP under strict power constraints, while the ILP serves more SU UEs than equal rate under strict interference constraints.

In the future work, we would like to develop the algorithm under multipath channels, where the feasibility of the transceiver beamforming depends on the angles of individual paths in the channel at SU BS. Therefore, the channel conditions, and hence the feasibility, should be checked before the beamforming step. 

\vspace{-3mm}

\bibliographystyle{IEEEtran}
\bibliography{IEEEabrv,references}

\begin{thebibliography}{1}
\providecommand{\url}[1]{#1}
\csname url@samestyle\endcsname
\providecommand{\newblock}{\relax}
\providecommand{\bibinfo}[2]{#2}
\providecommand{\BIBentrySTDinterwordspacing}{\spaceskip=0pt\relax}
\providecommand{\BIBentryALTinterwordstretchfactor}{4}
\providecommand{\BIBentryALTinterwordspacing}{\spaceskip=\fontdimen2\font plus
\BIBentryALTinterwordstretchfactor\fontdimen3\font minus
  \fontdimen4\font\relax}
\providecommand{\BIBforeignlanguage}[2]{{%
\expandafter\ifx\csname l@#1\endcsname\relax
\typeout{** WARNING: IEEEtran.bst: No hyphenation pattern has been}%
\typeout{** loaded for the language `#1'. Using the pattern for}%
\typeout{** the default language instead.}%
\else
\language=\csname l@#1\endcsname
\fi
#2}}
\providecommand{\BIBdecl}{\relax}
\BIBdecl

\bibitem{Goldsmith2012}
E.~Biglieri, A.~Goldsmith, L.~Greenstein, N.~Mandayam, and H.~Poor,
  \emph{Principles of Cognitive Radio}.\hskip 1em plus 0.5em minus 0.4em\relax
  Cambridge University Press, 2012.

\bibitem{Noam2012}
Y.~Noam and A.~Goldsmith, ``Exploiting spatial degrees of freedom in {MIMO}
  cognitive radio systems,'' in \emph{Communications ({ICC}), 2012 {IEEE}
  {International} {Conference} on}, Jun. 2012, pp. 3499--3504.

\bibitem{Tsinos2013}
C.~Tsinos and K.~Berberidis, ``Blind {Opportunistic} {Interference} {Alignment}
  in {MIMO} {Cognitive} {Radio} {Systems},'' vol.~3, no.~4, pp. 626--639, Dec.
  2013.

\bibitem{Xu2013}
T.~Xu, L.~Ma, and G.~Sternberg, ``Practical interference alignment and
  cancellation for {MIMO} underlay cognitive radio networks with multiple
  secondary users,'' Dec. 2013.

\bibitem{Du2012}
H.~Du, T.~Ratnarajah, M.~Pesavento, and C.~Papadias, ``Joint {Transceiver}
  {Beamforming} in {MIMO} {Cognitive} {Radio} {Network} {Via} {Second}-{Order}
  {Cone} {Programming},'' \emph{{IEEE} Trans. Signal Process.}, vol.~60, no.~2,
  pp. 781--792, Feb. 2012.

\bibitem{Du2013}
H.~Du and T.~Ratnarajah, ``Robust {Utility} {Maximization} and {Admission}
  {Control} for a {MIMO} {Cognitive} {Radio} {Network},'' \emph{{IEEE} Trans.
  Veh. Technol.}, vol.~62, no.~4, pp. 1707--1718, May 2013.

\bibitem{Ngo2014}
H.~Q. Ngo, E.~Larsson, and T.~Marzetta, ``Aspects of favorable propagation in
  {Massive} {MIMO},'' in \emph{Signal {Processing} {Conference} ({EUSIPCO}),
  2014 {Proceedings} of the 22nd {European}}, Sep. 2014, pp. 76--80.

\bibitem{Wang2011}
J.~Wang, P.~Urriza, Y.~Han, and D.~Cabric, ``Weighted centroid localization
  algorithm: Theoretical analysis and distributed implementation,''
  \emph{{IEEE} Trans. Wireless Commun.}, vol.~10, no.~10, pp. 3403--3413, 2011.

\bibitem{Friedlander1989}
B.~Friedlander and B.~Porat, ``Performance analysis of a null-steering
  algorithm based on direction-of-arrival estimation,'' \emph{Acoustics, Speech
  and Signal Processing, IEEE Transactions on}, vol.~37, no.~4, pp. 461--466,
  April 1989.

\end{thebibliography}


\end{document}